\documentclass[twocolumn,english,prb,showpacs]{revtex4}
\usepackage[T1]{fontenc}
\usepackage[latin9]{inputenc}
\usepackage{graphicx}
\usepackage{amssymb}

\usepackage{mathrsfs}

\usepackage{float}

\usepackage{indentfirst}

\makeatletter

\def\journal #1, #2, #3, 1#4#5#6{{\sl #1~}{\bf #2}, #3 (1#4#5#6) }

\usepackage{mathrsfs}
\usepackage{float}
\usepackage{indentfirst}

\def\eqa{
\begin{eqnarray}
    }
    \def\eea{
\end{eqnarray}
}
\newcommand{\eq}{
\begin{equation}
    }
    \newcommand{\ee}{
\end{equation}
}
\newcommand{\nn}{\nonumber\\}

\newcommand{\Tr}{{\rm Tr}}

\newcommand{\ua}{\uparrow}
\newcommand{\da}{\downarrow}

\makeatother

\usepackage{babel}

\begin{document}

\title{An Impurity Solver Using the Time-Dependent Variational Matrix Product State Approach}

\author{Lei Wang$^{1}$, Jia-Ning Zhuang$^{1}$, Xi Dai$^{1}$, and X. C. Xie$^{2,1}$}

\affiliation{$^{1}$Beijing National Lab for Condensed Matter Physics and Institute of Physics, Chinese Academy of Sciences, Beijing 100190, China }

\affiliation{$^{2}$Department of Physics, Oklahoma State University, Stillwater, Oklahoma 74078, USA}
\begin{abstract}
    We use the time dependent variational matrix product state (tVMPS) approach to investigate the dynamical properties of the single impurity
    Anderson model (SIAM). Under the Jordan-Wigner transformation, the SIAM is reformulated into two spin-1/2 XY chains with local magnetic fields along the $z$-axis.
    The chains are connected by the longitudinal Ising coupling at the end points. The ground state of the model is searched variationally within the space spanned by
    the matrix product state (MPS). The temporal Green's functions are calculated both by the imaginary and real time evolutions, from which the spectral information
    can be extracted. The possibility of using the tVMPS approach as an impurity solver for the dynamical mean field theory is also addressed. Finite temperature density
    operator is obtained by the ancilla approach. The results are compared to those from the Lanczos and the Hirsch-Fye quantum Monte-Carlo methods.
\end{abstract}

\pacs{71.27.+a, 71.10.Fd, 71.20.Be}

\maketitle
\section{Introduction} During the past twenty years, the dynamical mean field theory (DMFT) \cite{Metzner:1987p6797, Jarrell:1992p2251} has been quickly developed into a powerful method to solve the strongly correlated models on the lattice (for a review see \citet{Georges:1996p5571}). DMFT maps the lattice models to the corresponding quantum impurity models subject to the self-consistency conditions. Unlike the normal static mean field approaches, DMFT keeps the full local dynamics induced by the local interactions. DMFT has been successfully applied to various correlation problems, such as the Mott transition in the Hubbard model \cite{Georges:1996p5571, Bulla:2001p551} and the heavy fermion systems\cite{Si:2001p51, Gegenwart:2008p1203}.

In DMFT one encounters the problem of how to efficiently solve the
quantum impurity problems with self-consistently determined bath.
The impurity solver can be regarded as the \textit{engine} of DMFT,
which influences the efficiency and accuracy of DMFT calculations.
Since the invention of DMFT, many impurity solvers have been
developed. With the development of the modern computers, the
essentially exact numerical methods have received much attentions.
Among them the most used methods include the exact diagonalization
(ED) \cite{Caffarel:1994p1304}, Hirsch-Fye Quantum Monte Carlo
(HFQMC) \cite{Hirsch:1986p2060} and the numerical renormalization
group (NRG) \cite{Bulla:1998p597}. Most recently the continuous-time
quantum Monte Carlo (CTQMC) solver \cite{Werner:2006p2175,
Rubtsov:2005p2260} has been introduced. These methods can be divided
into two classes. ED and NRG \cite{Bulla:2007p2188} approaches are
based on the Hamiltonains, while the QMC solvers are based on the
Lagrangians (action). Generally the Lagrangian approach is favored
since the DMFT theories itself is derived in the Lagrangian
representation, and thus in the self-consistent process there is no
need to map the continuous hybridization function to the discrete
Hamiltonians. However the Hamiltonian approaches have the merit that
they work well at low temperatures and for real frequency which is
more relevant to the experimental quantities, since most of the
novel quantum phenomena in condensed matter physics happen at very
low temperature. By the term of "impurity solver", one means not
only to compute the ground state but also the whole spectral
functions, which include the lower energy quasi-particle parts as
well as the higher energy Hubbard bands in general. NRG approach
could resolve exponentially small scales at the expense of the
accuracy at intermediate and high energies.

The extensions of DMFT also call for the development of impurity
solvers. In recent years, LDA+DMFT has been developed very quickly
and successfully applied to many systems \cite{Anisimov:1997p2087},
see \citet{Kotliar:2006p5576} and \citet{Held:2007p4809} for the
reviews of the recent developments and applications. Since in the
real materials there are usually more than one orbitals involved,
therefore one needs to solve the quantum impurity problems with
multi-orbitals efficiently. A second direction of the development is
to study not the impurities but small clusters embedded in bath to
capture the spatial fluctuations\cite{Maier:2005p5551}. Therefor one
needs an efficient method to calculate the spectral properties of
smaller clusters. The requirement of solving complex impurity
problems (multi-orbital or cluster) puts hurdles on the ED and NRG
approaches.

A third direction of extension of the DMFT is to deal with the non-equilibrium systems \cite{Freericks:2006p6866}, which do not have the translational symmetry in the time domain. It requires the solver being able to work in the time domain, while to our knowledge most of the solvers work in the frequency (imaginary or real) domain. A recent attempt is made in the CTQMC approach \cite{Schmidt:2008p6350}. While this approach can handle arbitrary interaction strengths, it suffers from a dynamical sign problem which becomes severe at long time or for large bandwidth.

Therefore it is urgent to develop an impurity solver working at zero
temperature, which satisfies the following criteria. i) It can
capture both the low energy quasi-particle physics and the high
energy Hubbard bands. ii) It is easy to be generalized to
multi-orbital or cluster cases. iii) It works with real frequency
and gives the real time dynamical properties directly.

In this paper, we develop an impurity solver satisfying all these criteria based on the time-dependent variational matrix product states (tVMPS) approach. The variational approach directly attacks the strongly correlated problems by an educated guess of the many body wavefuncitions. By introducing more variational parameters one could enlarge the dimensions of the variational space but it makes the variation process harder in general. For example, in the case of Gutzwiller variational wave function, the evaluation of the expectation value is often done approximately by introducing the Gutzwiller approximation \cite{Gutzwiller:1965p780, Vollhardt:1984p5546}. Matrix product states (MPS) are states where the coefficients of the wave function are a product of matrices depending on the local lattice site states. They are generated naturally from the NRG and DMRG calculations. Actually the latter approach can be reformulated into the variational approach within the space spanned by the MPS \cite{Verstraete:1999p5871}. Time evolution algorithm \cite{Vidal:2004p5978} based on MPS was first proposed from the quantum information perspective, and then been translated into the language more access to the many-body theorists \cite{White:2004p5977, Daley:2004p5979}, see \citet{Verstraete:2008p5609, GarciaRipoll:2006p5994} and \citet{Schollwoeck:2006p1926} for reviews of the time evolution algorithm of the MPS.

There are some previous attempts using the VMPS approach to study
the quantum impurity problems. \citet{Weichselbaum:2005p5639}
studied the spectral property of SIAM in the presence of a magnetic
field using the correction vector approach, which is first proposed
in the context of dynamical density matrix renormalization group
(DDMRG) \cite{Jeckelmann:2002p5997}. \citet{Saberi:2008p5625}
performed detailed comparison of the VMPS and the NRG approach on
the SIAM. \citet{Holzner:2008p5660} used the VMPS formalism to study
the two-lead, multi-level Anderson impurity model. Along the line of
developing fast impurity solvers, several of the authors have
developed the Gutzwiller based impurity solver
\cite{Zhuang:2009p6335}, which is suited for combining with the LDA
and studying the low temperature properties of multi-orbital models.
There were previous attempt of using the DMRG as the impurity solver
of the DMFT. \cite{Garcia:2004p6795, Nishimoto:2004p6894}

The organization of this paper is as follows. In Sec. \ref{sec:gs}
we give an overview of the computation of the ground state of the
SIAM in the VMPS formalism, where an unfold \cite{Saberi:2008p5625}
technique greatly reduces the computational effort. In Sec.
\ref{sec:evolution} we describe the algorithm for time evolution in
both real and imaginary times, by which one could calculate the
Green's functions in the time domain. In Sec.\ref{sec:ft} we
describe the fit and extrapolate scheme to extract the spectral
function from the real time data.  In Sec. \ref{sec:dmft} we address
the DMFT self-consistent loop where a fitting procedure from the
continuous hybridization function to the SIAM chain Hamiltonian is
needed. Finally, we conclude the paper by making remarks on the pros
and cons of the present solver and make an outlook for future
developments. In the appendix we  generalize the solver to finite
temperature, where ancillary sites which play the role of the heat
bath are introduced.

\section{VMPS approach to the Ground State of the SIAM \label{sec:gs}}

\begin{figure}
    [tbp] \centering
    \includegraphics[height=6cm, width=8cm]{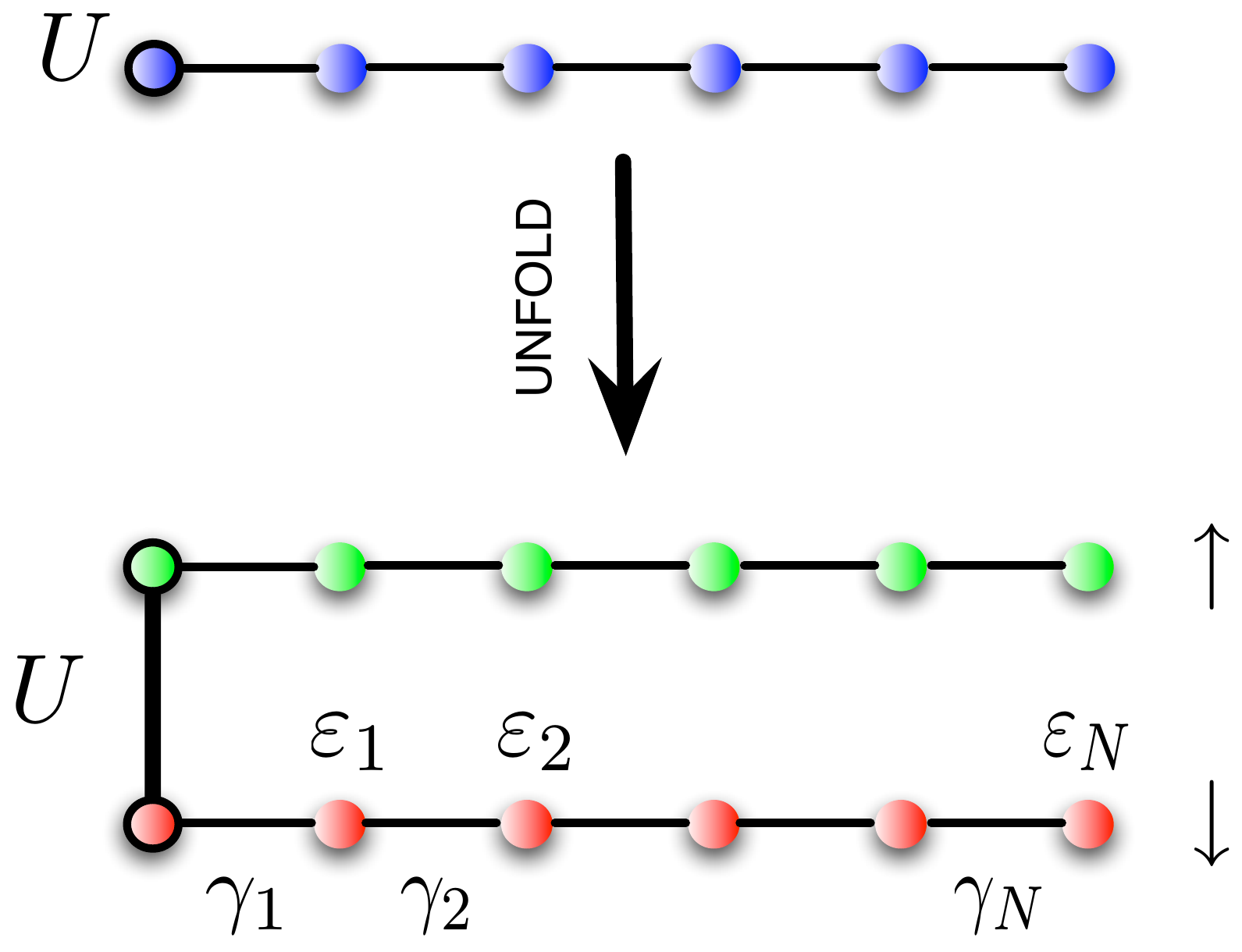} \caption{(Color online) Unfold the SIAM chain. The SIAM chain is separated into two parts with different spins, they are connected at the leftmost end where the bold bond denotes the Hubbard interaction $U$.} \label{fig:chain}
\end{figure}

The action of the single impurity Anderson model (SIAM) is
\begin{eqnarray}
    S_{SIAM} &=&\int_{0}^{\beta} d\tau \int_{0}^{\beta} d\tau^{\prime} c_{\sigma}^{\dagger}(\tau) [(
    \partial_{\tau} -\mu )\delta_{\tau,\tau^{\prime}} +\Delta(\tau,\tau^\prime)]c_{\sigma}(\tau^\prime) \nn & & + \int_{0}^{\beta} d \tau U n_{\ua}n_{\da}
\end{eqnarray}
It is a zero dimensional problem, and there only exists the fluctuation in the temporal axis, which is captured by the time dependent hybridization function $\Delta(\tau, \tau^\prime)$. One could de-integral the action by introducing the noninteracting bath degree of freedoms. The electron hops into the bath, travels in it for a while and then hops back to the impurity site thus brings in the temporal nonlocal correlations. The de-integral process is not unique, and in the DMFT context, both the star and the chain geometry of the bath degree of freedom have been studied. In the present study we reformulate the SIAM into a one dimensional chain with the nearest neighbor hopping, see Figure~\ref{fig:chain}. In the framework of DMFT the Hamiltonian parameters of the chain are determined self-consistently, and generally has no translational invariance. We deal with the chain with typical chain length of $10$ to $20$ sites in the present study.

By introducing the Jordan-Wigner transformation (JWT) for the spin up and down fermions separately, $s^-_i (\tau^{-}_i) = \left( \prod_{k < i} p_{k\sigma} \right) c_{i\sigma}$ where $p_{k\sigma} = (- 1)^{n_{k\sigma}} $ is the JW sign, one could unfold the SIAM chain into two spin-$\frac{1}{2}$ $ XY$ chains. They are coupled at the end point by the longitude Ising coupling due to the Comlomb interaction. The onsite energy of each site is mapped into the local magnetic filed along the $z$-direction. Since the hopping is between the nearest neighbors, the JWT sign does not shows up in the calculation of the ground state. A simple division of the Hamiltonian into terms acting on the odd and even bonds are possible, see Sec.\ref{sec:evolution}.

In terms of $1/2$ spins, the Hamiltonian is

\begin{eqnarray}
    H_{SIAM} & = &H_{imp} + H_{bath} + H_{hyb} \nn \nn
    H_{imp} & = & - \mu(s^+_0 s_0^- +\tau^+_0 \tau_0^-) + U (s^+_0 s_0^-) (\tau^+_0 \tau_0^-) \nn
    H_{bath} & = & \sum_{i=1}^{N_{bath}} \varepsilon_i (s^+_i s^-_i + \tau^+_i \tau^-_i) \nn &+& \sum_{i=1}^{N_{bath}-1} \gamma_{i+1} (s^+_i s^-_{i + 1} + \tau^+_i \tau^-_{i + 1} + h.c.) \nn
    H_{hyb} & = &  \gamma_{1} (s^+_0 s^-_{1} + \tau^+_{0} \tau^-_{1} + h.c.) \nn
\end{eqnarray}

where $s_i^{\pm} = \frac{\sigma_x \pm i \sigma_y}{2}$ are the ladder operators for spin one half. $s$ and $\tau$ denote the spin up/ down fermion operators.
\begin{figure}
    [tbp] \centering
    \includegraphics[height=4cm, width=8cm]{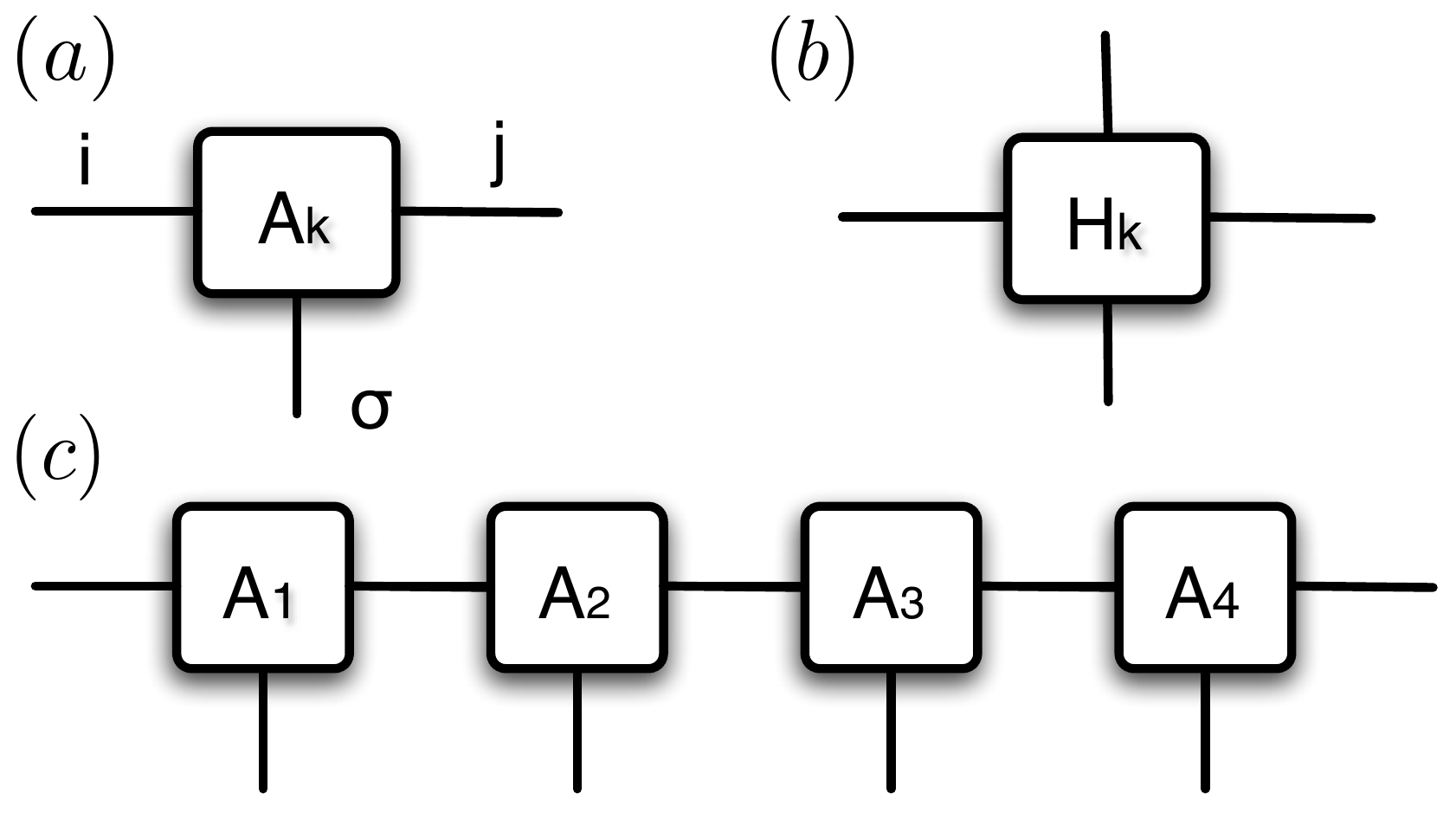} \caption{(Color online) (a). The $A_{ij}^{[\sigma_{k}]}$ has three indices. $\sigma_{k}$ is the physical index run from $1$ to $d=2$ for SIAM. $i$ ($j$) is the visual index runs from $1$ to $\chi_{k}$ ($\chi_{k+1}$). They control the precision of the VMPS calculation. (b). An operator acts on a single site (c). A MPS where all of the connected bonds have been contracted.} \label{fig:mps}
\end{figure}

\begin{figure}
    [tbp] \centering
    \includegraphics[height=12cm, width=8cm]{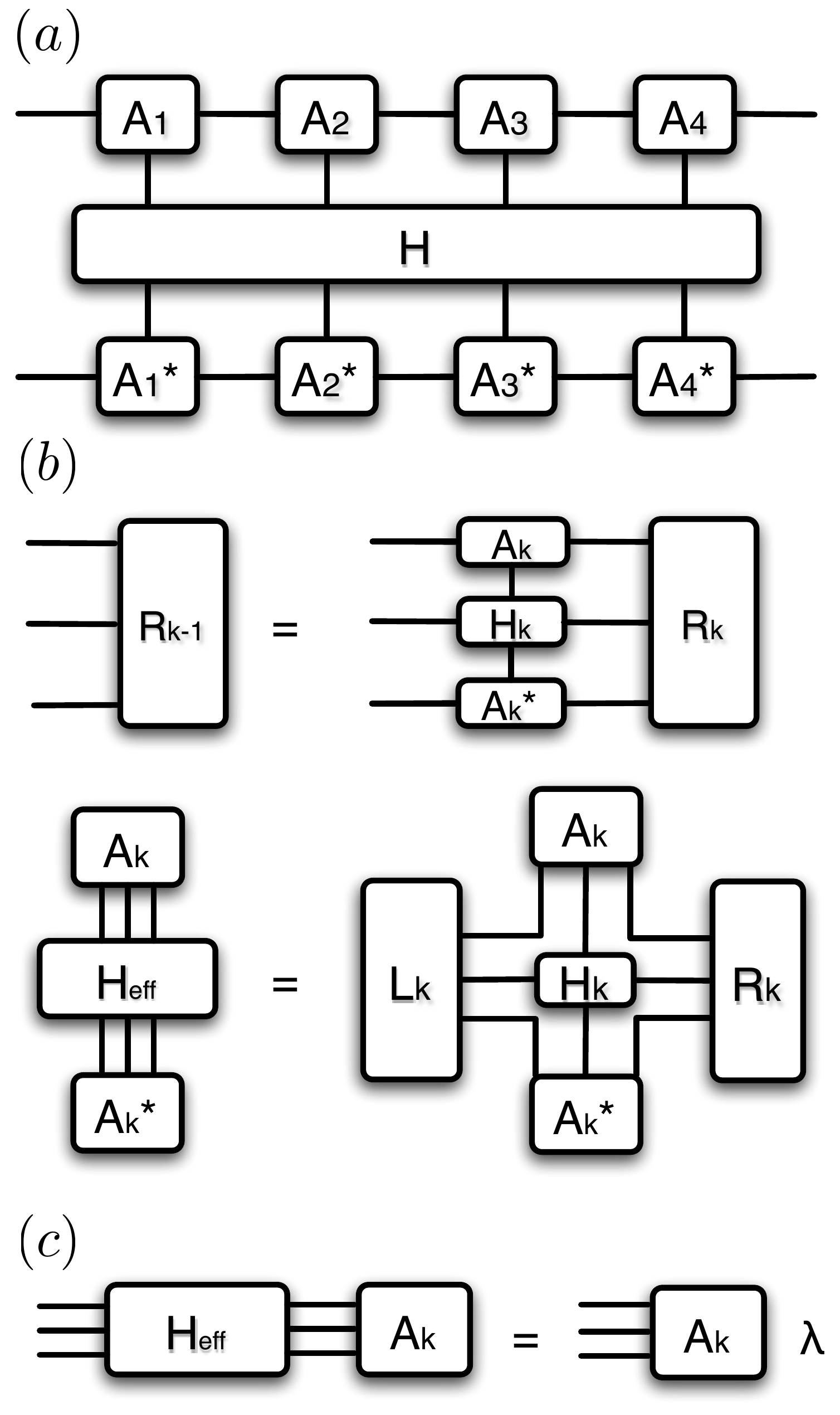} \caption{(Color online) (a). The expectation value of the operator $H$ over the MPS. (b). Tracing out all the indices on the sites other than $k$ gives the effective Hamiltonian $H_{eff}$ on the $k$-th site. (c). The minimization problem on site $k$ is equivalent to the eigen-problem of the effective Hamiltonian } \label{fig:mps_gs}
\end{figure}

The matrix product state (MPS) $|\psi_{MPS}\rangle = \sum_{\sigma_1,\sigma_2,...,\sigma_N}\Tr(A_{1}^{[\sigma_{1}]} A_{2}^{[\sigma_{2}]} ... A_{N}^{[\sigma_{N}]})|\sigma_{1},\sigma_2,...,\sigma_N \rangle $. $\sigma_k=1,...,d$ is the physical index, where $d=2$ is the dimension of the local Hilbert space. $A^{[\sigma_{k}]}$ is a matrix with the visual dimension $\chi_{k}\times \chi_{k+1}$. There is an efficient scheme based on the transfer matrix to handle them in low dimensions, \textit{i.e.} to calculate the overlap between two MPSs, and the expectation value of an operator over two MPSs, the operation of a local operator on a given MPS. They are shown schematically in Fig.\ref{fig:mps_gs} and see \citet{Verstraete:2008p5609} for a reference.

We minimize the expectation value $\langle \psi_{MPS}| H_{SIAM} |\psi_{MPS}\rangle$ within the space spanned by the normalized MPS. Such a problem is solved very efficiently by the alternating least squares scheme, in which we perform the optimization of the matrices site by site. In each step, we fix all but the matrix on the current site, see Figure~\ref{fig:mps_gs}. By contract all of these indices we define the effective Hamiltonian on the current site. The minimization problem becomes quadratic and is equivalent to an eigen-problem of the size $\chi_{k}\times \chi_{k+1}\times d$. Since only the ground state is needed, we solved it by the Lanczos method. The key step in the Lanczos step is the computation of the matrix-vector multiplication:
\[ \sum_{l,\sigma,r} H_{eff\{l^{\prime},\sigma^{\prime} ,r^{\prime}\}\{l,\sigma,r\}} A^{[\sigma]}_{l,r} = \sum_{m} h^{m}_{\{l^{\prime},\sigma^{\prime},r^{\prime}\}\{l,\sigma,r\}} A^{[\sigma]}_{l,r} \]
Where $H = \sum_{m} h^{m}$ can be parallelized over each term in the Hamiltonian.

\section{Time Evolution \label{sec:evolution}}

To calculate the retarded Green's function $G_{\sigma}^r (t) = - i \theta (t) \langle GS|\{c_{\sigma} (t), c_{\sigma}^{\dagger} \}|GS\rangle $, one first applies the $c^{\dagger}_{\sigma}$ or $c_{\sigma}$ on the ground state from the previous VMPS calculations: $| \phi(0) \rangle = c_{\sigma}^{\dagger} | GS \rangle$, $| \chi(0) \rangle = c_{\sigma}| GS \rangle$. Then let states evolute in real time to get $| \phi (t) \rangle = e^{- i (H - E_G) t} c^{\dagger}_{\sigma} |GS \rangle$ and $| \chi (t) \rangle = e^{i (H - E_G) t} c_{\sigma}|GS \rangle$. Thus the retarded Green's function can be calculated as
\begin{equation}
    G^r_{\sigma} (t) = - i \theta (t) [\langle \phi(0) | \phi(t) \rangle + \langle \chi(0) | \chi (t) \rangle]
\end{equation}

The operation of the local operators on the ground state MPS is
\begin{equation}
    c^{\dagger}_{k\ua}| GS \rangle = \prod_{l<k} (p_{l\ua})_{\sigma_l\sigma_l^{\prime}}A^{[\sigma_l^{\prime}]} (s^{+}_{k})_{\sigma_k\sigma_k^{\prime}} A^{[\sigma_k^{\prime}]} \prod_{m>k} A^{[\sigma_m]} | \vec{\sigma}_l,\sigma_k,\vec{\sigma}_m\rangle \label{eqn:cdagger_gs}
\end{equation}
Where a string of JW signs and a $d\times d$ matrix $s^{+}$ act on the $k$-th site. After the multiplication, the new state is still represented as the the MPS of the same rank.

The real-time evolution is performed by first splitting the evolution operator into small pieces $e^{-iHt}=(e^{-iH\Delta t} )^M$, where $\Delta t=\frac{t}{M}$, then applying the Trotter decomposition. To the second order one has $e^{-iH{\Delta}t}=e^{-iH_{o}{\Delta}t/2} e^{-iH_{e}{\Delta}t/2} e^{-iH_{Z}{\Delta}t}e^{-iH_{e}{\Delta}t/2} e^{-iH_{o}{\Delta}t/2} + O((\Delta t)^3)$. Where $H=H_{o}+H_{e}+H_{Z}$. $H_e$ and $H_o$ are two particle terms act on even and odd bonds and $H_Z$ includes the single particle terms. Since each term within $H_{\alpha}$, ($\alpha=e,o,Z$) commutes one could apply the evolution operator $U_{\alpha}(\Delta t)=e^{-iH_\alpha\Delta t}$ associate with each bond one after another without leading to further errors. In the particle-hole symmetric case, the onsite energies of the chain are zero and the chemical potential $\mu=U/2$ can be put into the interaction term. Thus the single particle evolution operator $U_{Z}=1$

\begin{figure}
    [tbp] \centering
    \includegraphics[height=6cm, width=8cm]{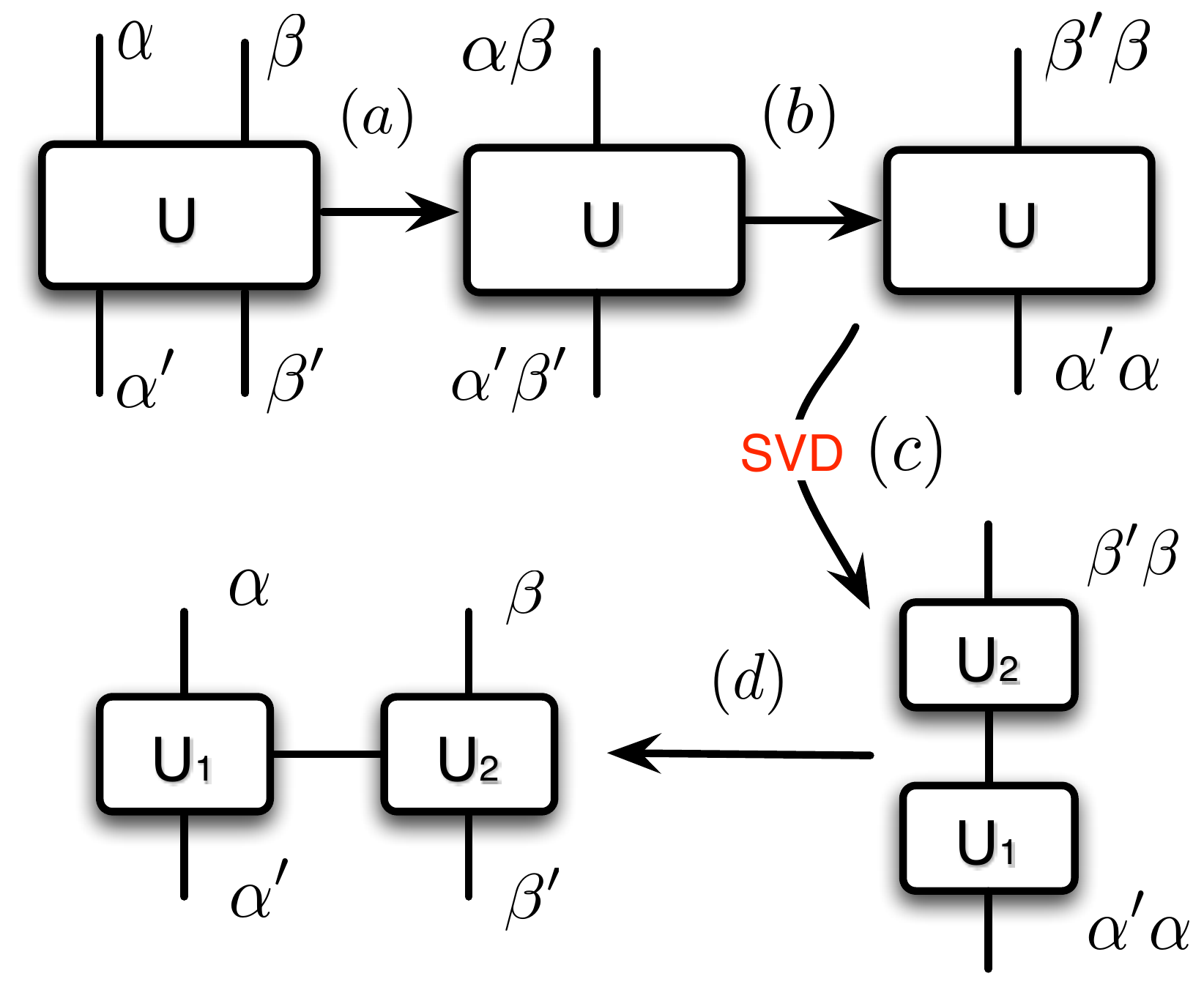} \caption{(Color online) Writing the two particle evolution operator into a MPO. (a). The physical indices of the four indices tensor are merged, giveing a $d^2\times d^2$ matrix. (b). The order of the indices are exchanged. (c). SVD of the matrix separates it into two parts (d). Rearrangement of the physical indices gives the MPO representation of the two-sites evolution operator. } \label{fig:mpouv}
\end{figure}

The operation of the single site evolution operator $U_{Z}$ is similar to the applying of the creation/annihilation operators, Eqn~\ref{eqn:cdagger_gs}. The operation of the two site evolution operators $U_{e}$ or $U_{o}$ is shown schematically in Fig.\ref{fig:mpouv}. The physical indices for the two sites are first merged and exchanged, then the $d^2\times d^2$ operator is separated into two matrices $U_{1}$ and $U_{2}$ by the singular value decomposition (SVD). Each of them only carries the physical indices of one site. In both cases, the evolution operators are written as the matrix product operator (MPO) \cite{Murg:2008p6110}, which has four indices, two for the visual dimensions and two other for the physical indices, see Fig.\ref{fig:mps}. Generally after applying them onto the target states, the resulting states could not be written as the MPS with same visual dimension as before. Thus one needs to project the state into the original MPS space. It is done variationally by minimizing the norm: $ \mathcal{N} = \| | \phi^{\prime} \rangle - U_{\alpha}(\Delta t)| \phi (t) \rangle \|$, and approximately one has |$\phi(t+{\Delta}t)\rangle=|\phi^{\prime}\rangle$. The minimization procedure can be carried out following the similar procedures as in Sec.\ref{sec:gs}. It is performed every step after the applying of the evolution operator on the original state. In the course of evolution, we increase the bond dimension to maintain the smallest eigenvalue of the reduced density matrix larger than a threshold $w_{min}=10^{-10}$.

\begin{figure}
    [tbp] \centering
    \includegraphics[height=6cm, width=8cm]{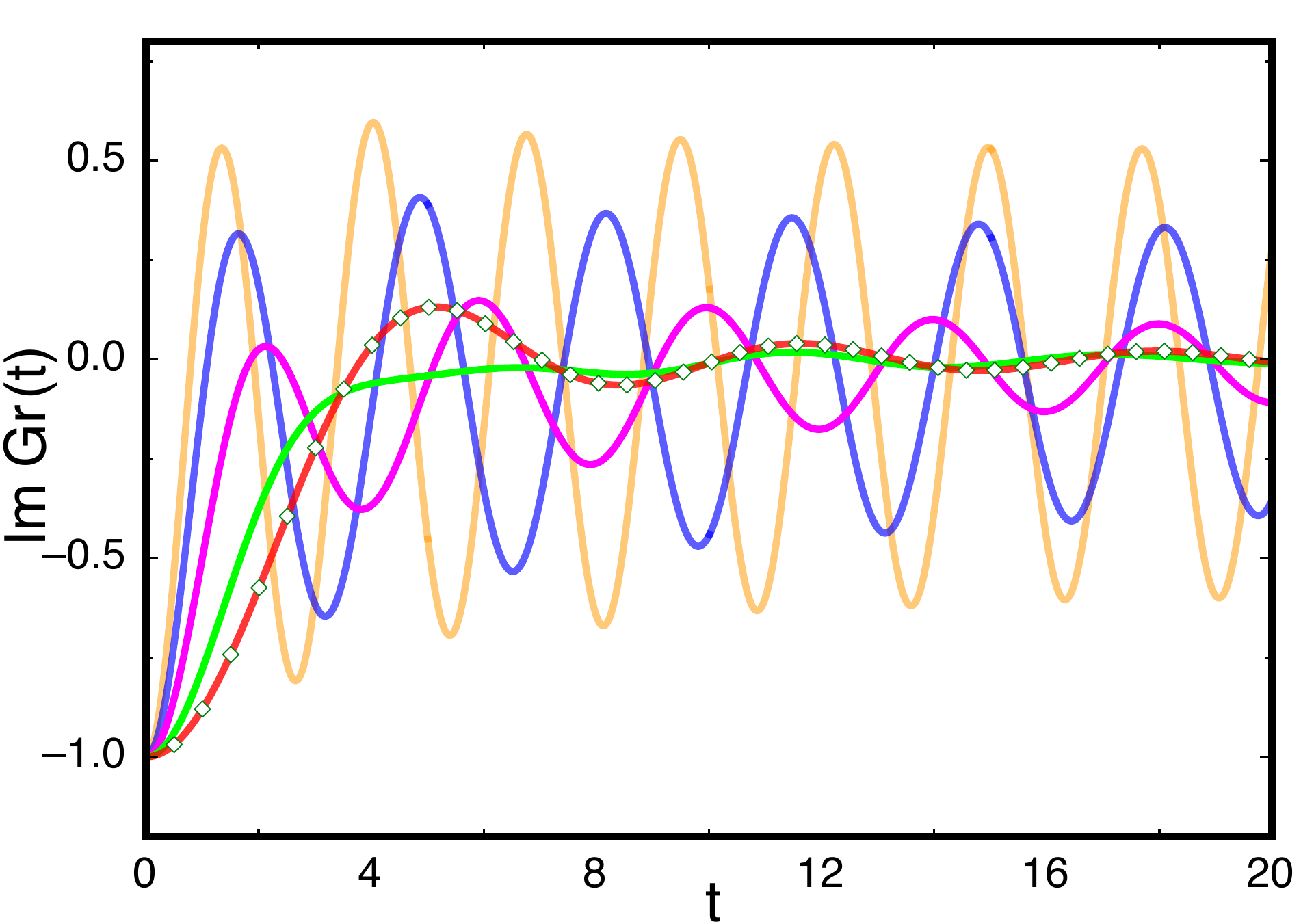}

    \caption{(Color online) The zero temperature retarded Green's function $G^r(t)$ for a $20$-site SIAM chain. Where $\varepsilon_{n}=0, \gamma_{n}=0.5$, the interacting strength $U=0,1,2,3,4,5$ from the bottom to top. The diamond shaped dots indicate the analytical result for $U=0$. Finite size effect of the chain does not show up for this time scale, so the Fourier transform of the real time Green's function gives spectral functions. }

    \label{fig:gt}
\end{figure}

See Figure~\ref{fig:gt} for the real time Green's function for different interacting strength. The retarded Green's function $G^r(t)$ evolutes from the Bessel function at the noninteracting limit to the cosine function at the atomic limit. In between one has the oscillated decaying curves, and the period of the oscillation gives the frequency of the Hubbard band. The normalization of the density of states (DOS) is automatically fulfilled since it implies that $G^{r}(t=0)=-1$. There may be a concern on the orthogonality catastrophe (OC)\cite{Anderson:1967p6864,Anderson:1967p6865} which states that in the thermodynamic limit local perturbation leads to complete reconstruction of the ground state of a fermionic system in such a way that the overlap of the "old" and "new" ground-state wave functions is proportional to $N^{-\alpha}$. But since we are dealing with finite $N$ here, the OC is irrelevant. \cite{Anderson:1967p6864,Anderson:1967p6865}. By comparing the real time Green's function with different bath sites, we notice that even with small number of sites, one could reproduce the resulting spectral information for the thermal-dynamical limit, as long as the wavefront created by the adding (removing) an electron on the impurity site does not reach the boundary.

Imaginary time Green's function could also be calculated similarly, in which we perform the evolution algorithm along the imaginary time axis. Imaginary time evolution technique could also be used to search the ground state. The imaginary time Green's function $G(\tau)$ is a decayed function from $n_{\sigma}-1$, and it is decaying faster for larger interacting strength $U$. For the particle-hole symmetric case, the decaying form approaches single mode decay $-\frac{1}{2}e^{-\mu\tau}$ in the atomic limit.

\section{Extract Spectral Function \label{sec:ft}}

\begin{figure}
    [tbp] \centering
     \includegraphics[height=6cm, width=8cm]{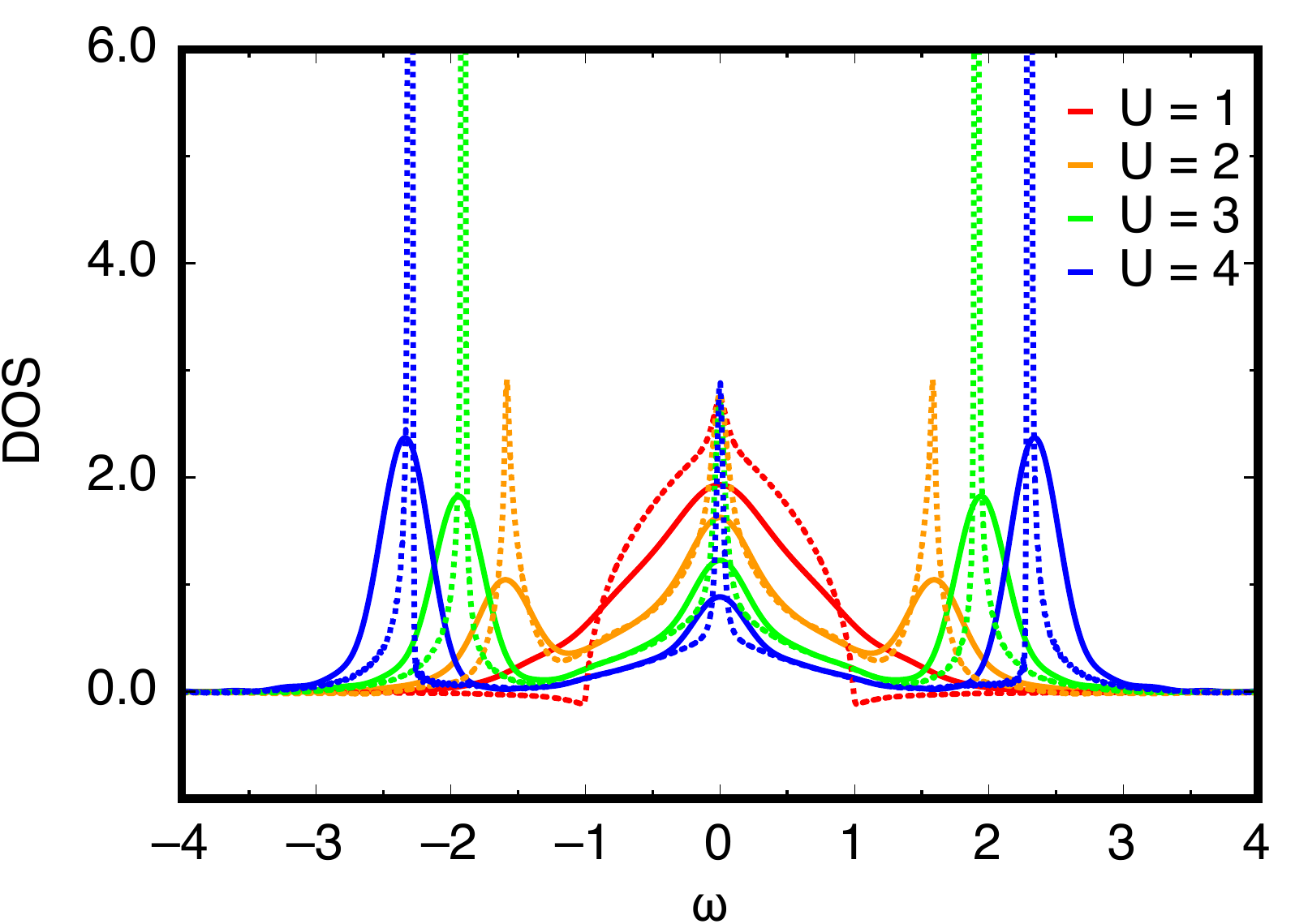}
     \caption{(Color online) DOS of a SIAM for $U=1,2,3,4$ from directly FT (full line) of real time data (Fig.\ref{fig:gt}) and fit the long time behavior and then perform FT (dash line).} \label{fig:fit}
\end{figure}

The zero temperature spectral information can be extracted from the imaginary time Green's function $G(\tau)$, by the maximum entropy continuation similar to the procedures in the PQMC approach to the SIAM\cite{Feldbacher:2004p6858}. Or one could get the spectral information from the Fourier transform of real time Green's function $G^{r}(t)$. \cite{Barthel:2009p6149, White:2008p6155}.

In the Fourier transformation (FT) approach, to avoid the negative DOS, generally one multiplies the time domain data with a window function ($W(t)=e^{-4(\frac{t}{T_{max}})^2}$ is used here) which is a decayed function from zero time to the large time cutoff, but this method inevitably broaden the peak in the spectral function and also drops a large amount of raw data. Generally there are two approaches to avoid the overboadening: the linear prediction (LP) and the fitting and extrapolate (FE) procedure \cite{White:2008p6155, Barthel:2009p6149}. We leave the LP approach for further investigation and adopt the FE approach which we believe is more under control.

We fit parts of the raw data with $A\cos(\omega (t-t_0))/t^{\alpha}+Be^{-\beta t}$, where $A,B,t_0, \alpha, \beta$ are fitting parameters. Then the function is extrapolate to very large time. By this method we have taken use of most of the raw data. A comparison of DOS from directly FT and after "fit and extrapolate" is shown in Fig.\ref{fig:fit}. Even from directly FT data one could recognize zero component Kondo peak and the development of Hubbard band as $U$ increases. However, the Fridel sum rule which states the pinning of the $A(0)=\frac{2}{\pi D}$ is violated. Actually, the conservation of zero frequency peak relies on the conservation of the areas enclosed by the $G^r(t)$ curve. This means the resolution of the Kondo peak relies on very accurate long time behavior of the Green's function. By the FE procedure before FT, this conservation could be fulfilled, Fig.\ref{fig:fit}. However, the fitting ansatz inevitably introduces prior knowledge about the spectral and gives much sharper peaks in present case. Note that causality is violated (negative dos) in $U=1$ case.

We find it is more stable to extract the spectral function from the $G(\tau)$ by maximum entropy method (MEM). Since the data is free from statistical error, the procedure is more stable and less involved than the continuation of Monte Carlo results. As stated in early work of Naef \textit{et al} \cite{Naef:1999p12433}, although the continuation approach may be still insufficient for resolving specific line shapes, it is appropriate for identifying gaps and peak positions, which is more relevant in DMFT studies. So in present paper MEM is still adopted to get DMFT converged spectral functions.



\section{DMFT Self-consistent \label{sec:dmft}}

\begin{figure}
    [tbp] \centering
    \includegraphics[height=8cm, width=8cm]{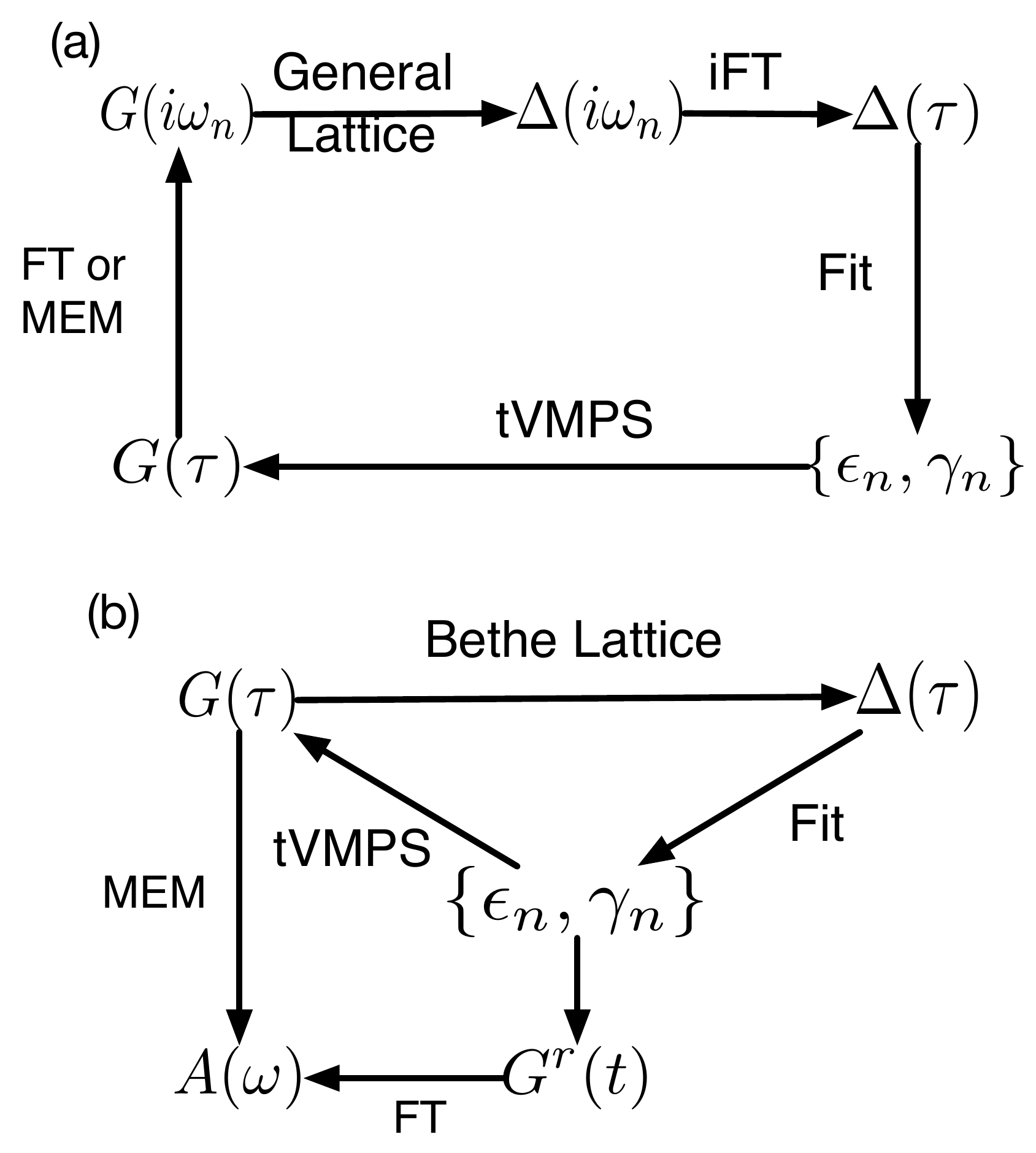}
    \caption{Flow diagram of DMFT self-consistent calculations. tVMPS solves Green's function for a SIAM chain Hamiltonian in temporal domain. Self-consistent process gives new hybridization function $\Delta(\tau)$, from which the Hamiltonian parameters $\{\varepsilon_i, \gamma_i \}$ are fitted, thus closes the loop. (a). For general lattice, one performs Fourier transform or analytical continuation to translate $G(\tau)$ to frequency domain, and then determines new hybridization function, information of non-interaction DOS enters in this step. (b). The self-consistent loop simplifies for infinite dimension Bethe lattice, where Matsubara Green's function $G(\tau)$ directly determines the new hybridization function. Spectral function $A(\omega)$ could be gotten by MEM continuation of the converged Matsubara Green's function $G(\tau)$ or FT of retarded Green's function $G^{r}(t)$, in the latter case, real time evolution using converged Hamiltonian parameters is firstly performed. }
    \label{fig:dmft_loop}
\end{figure}

After getting the temporal Green's function, one has essentially solved the impurity problem. One should plug it into the DMFT self-consistent loop to produce the SIAM chain Hamiltonian for the next iteration step. For demonstration purpose we consider the case of the Bethe lattice where the self-consistent procedure is greatly simplified. By multiplying $\frac{D^2}{4}$ with $G(\tau)$ one gets the hybridization functions $\Delta(\tau)$, where $D$ is the half-bandwith of the semi-ellipse DOS. For the general lattice, the self-consistent loop should be 1) Using Fourier transform or MEM to get the Green's function in the frequency domain, and do the self-consistent following the conventional routine. This kind of self-consistent procedure was adopted in PQMC solver\cite{Feldbacher:2004p6858}.  
or 2) Just do all of the computation in the time domain. In that case, one needs to do inverse of the Green's function matrix in time domain and convolution of  non-interacting DOS with each matrix element. It is a formidable task but still could be done, as one encountered in the non-equilibrium DMFT. \cite{Freericks:2006p6866}

The key problem then is how to determine the Hamiltonian parameters
of the SIAM chain from the continuous hybridization function
$\Delta(\tau)$. Similar problem is encountered in other Hamiltonian
based solvers such as ED, NRG and DDMRG approaches. In the ED
approach the step is determined by the conjugated gradient
minimization \cite{Georges:1996p5571} or continued fraction
expansion \cite{Si:1994p56} of the Green's function in the frequency
domain, but due to the finite size of the effective bath that can be
dealt with, one has to truncate and leads to errors in this step. In
the NRG approach the logarithmic discretization procedure is adopted
\cite{Bulla:2008p6336,Bulla:1998p597}, while in the DDMRG approach,
no logarithmic energy separation is assumed, so a direct
tri-diagonalization scheme can be used, \textit{i.e.} first fit the
hybridization function with a star shaped Hamiltonian, then using
Lanczos method to tri-diagonalize the Hamiltonian, and the
diagonal/off diagonal matrix elements are the onsite/hopping
parameters for a chain Hamiltonian.

In the present case, we fit $\Delta(\tau)$ with the Hamiltonian parameters of the SIAM chain by minimizing
\begin{equation}
\chi_{\Delta}^2=\sum_{i=1}^{N_{\tau}} |\Delta(\tau_i)-\tilde{\Delta} (\tau_i)|^2
\end{equation}
$\tilde{\Delta}(\tau)$ is the hybridization function for a noninteracting SIAM chain with onsite energy $\varepsilon_i$ and hopping amplitude $\gamma_i$, $i=1,2,...N_{bath}$. $ \tilde{\Delta}(\tau) = \sum_{l}\gamma_1^2 |U_{1 l} |^2 e^{- E_l \tau} [\theta (\tau) (n_l - 1) + \theta(- \tau) n_l]$, where $U$ and $E$ are eigenstate and eigenvalue of the noninteracting Hamiltonian $H_{bath}$, satisfying $H_{bath}U=UE$. $n_l$ is the occupation number on level $l$, and $n_l =\theta(-E_l)$ for zero temperature. We use conjugate gradient method to search within the parameter space spanned by $\varepsilon_i$ and $\gamma_i$ which minimize $\chi_{\Delta}^2$. Generally there are $2N_{bath}$ parameters, while in the particle-hole symmetric case all of the $\varepsilon_n$ is kept as zero and therefore one has $N_{bath}$ fitting parameters. Since we are dealing with relatively larger number of bath sites comparing with ED approach and the function to be fitted is more well behaviored (different from the fitting $G(i\omega_{n})$ in frequency domain, one has $\Delta(\tau)$ which is more smooth and featureless), the fitting is more reliable. In the present study, the average fitting error can be reduced to $10^{-6}$ per data point.

\begin{figure}
    [tbp] \centering
    \includegraphics[height=7cm, width=8cm]{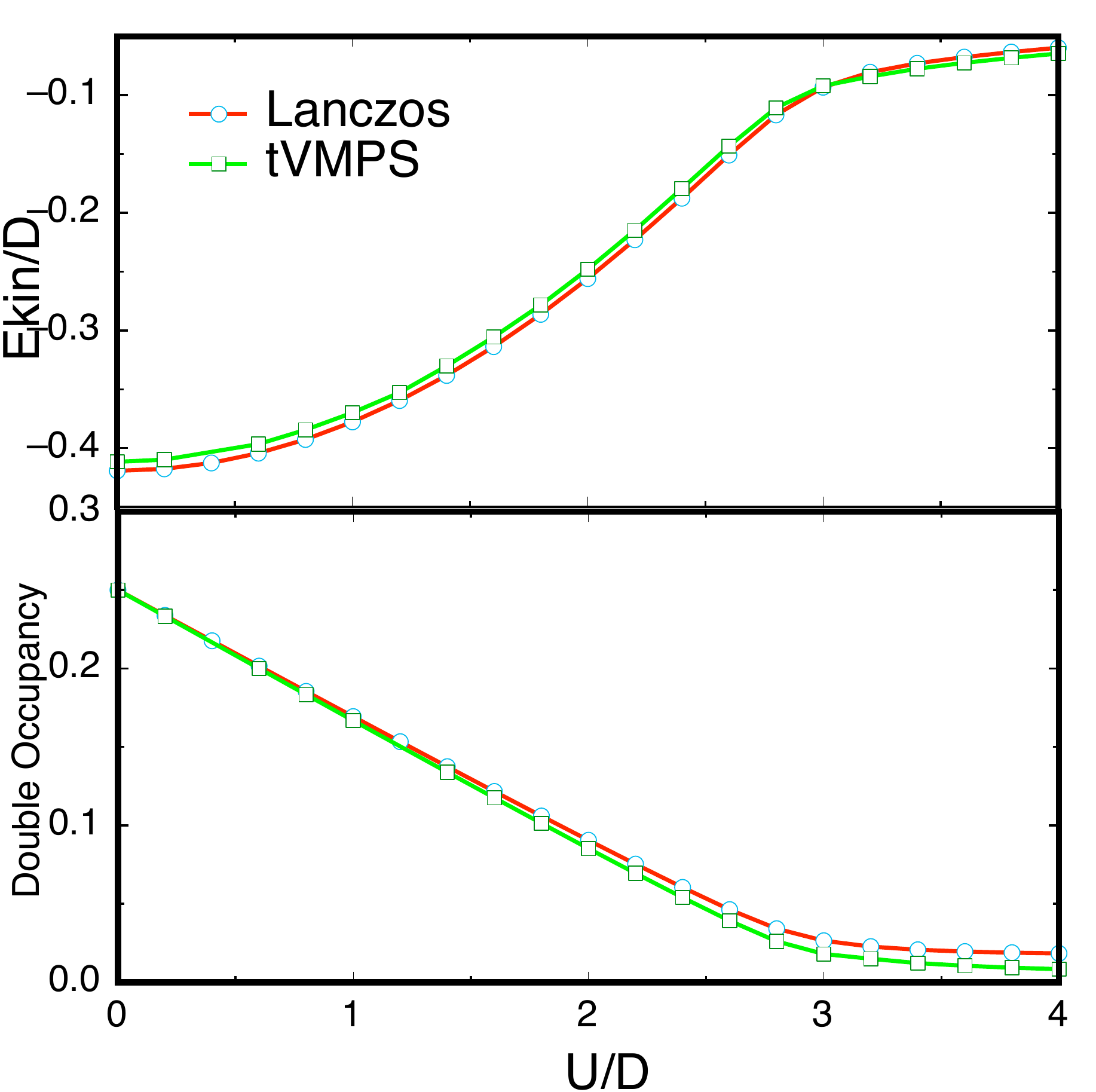} \caption{(Color online) Physical quantities at $T=0$ from the tVMPS solver compared with the Lanczos solver. Upper panel: The average kinetic energy per site, Lower panel: The average double occupancy per site versus the interacting strength $U/D$. } \label{fig:vmps_lanczos}
\end{figure}
\begin{figure}
    [tbp] \centering
     \includegraphics[height=6cm, width=8cm]{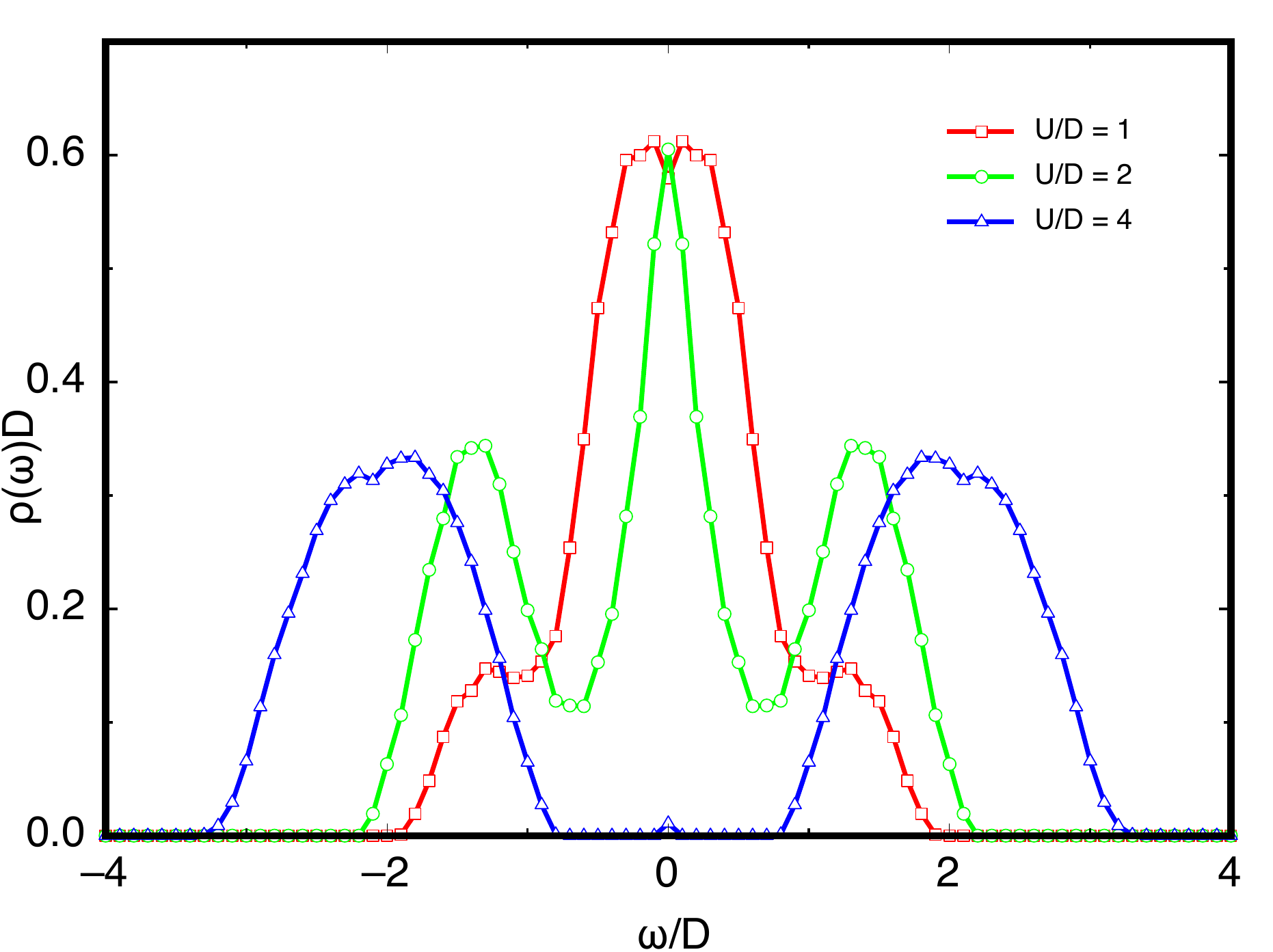}
     \caption{(Color online) The DOS extracted from the converged $G(\tau)$ by MEM.} \label{fig:mem}
\end{figure}

%
After the DMFT loop converges, one could get various information from the converged Green's function $G(\tau)$ and the MPS. {\textit e. g.} the kinetic energy and the double occupancy rate. See Fig.\ref{fig:vmps_lanczos} for the comparison of the results from the tVMPS and the Lanczos solver at zero temperature. The long time asymptotic behavior for the interaction strength $U$ larger than the critical $U_c$ shows the presence of the charge excitation gap. One could also recognize the metal-insulator transition clearly from the DOS on the Fermi surface $A(0)=-\lim_{\beta\rightarrow\infty} \frac{\beta}{2} G(\frac{\beta}{2})$ and the suppression of the double occupancy. The real frequency information can be extracted from the analytical continuation approaches, such as the Pade approximation or the MEM. The MEM inverted the integral transformation $-G(\tau)=\int_{-\infty}^{\infty}K(\omega,\tau) A(\omega) $, where symmetrical kernel $K(\omega,\tau)=\frac{1}{2}\frac{e^{-\omega\tau}+e^{(\tau-\beta)\omega}}{e^{-\beta\omega}+1}$ is used to persevere the particle hole symmetry $A(\omega)=A(-\omega)$, $\beta$ is a fictitious temperature $\beta=128$ in our calculation. The results are shown in Fig.\ref{fig:mem}.

\begin{figure}
    [tbp] \centering
    \includegraphics[height=6cm, width=8cm]{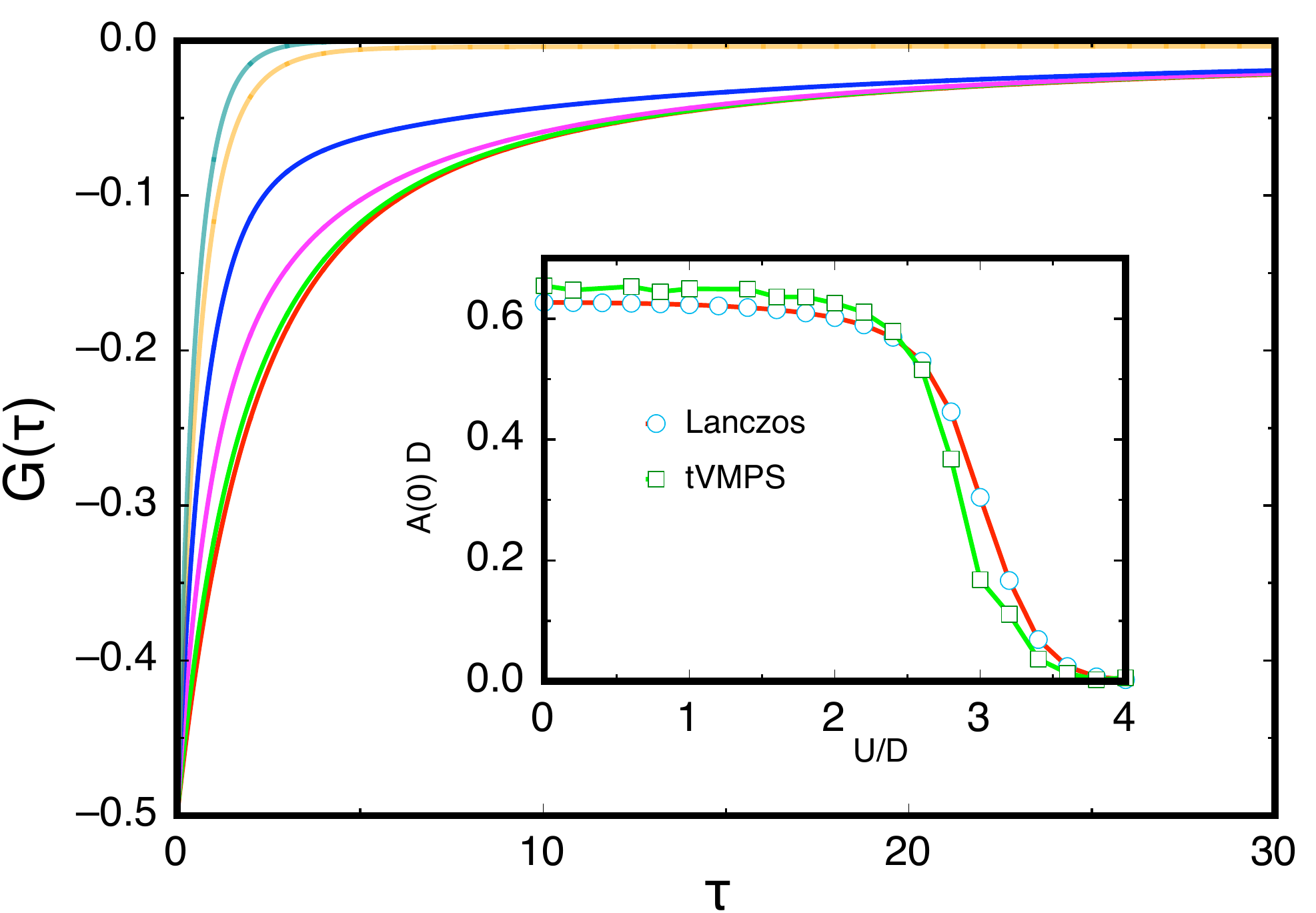} \caption{(Color online) Main figure: The DMFT converged zero temperature Green's functions by the tVMPS solver with $18$ sites. Where the interaction strength $U=0, 0.8D, 1.6D, 2.4D, 3.2D, 4.0D$ from bottom to the top. The long time asymptotic behavior of $G(\tau)$ differentiates the metallic and the Mott insulator states. In set: The density of states on the Fermi surface $A(0)$, which are compared with the results from a Lanczos solver with $8$-site. Slightly dropping of $A(0)$ in the metallic side is due to finite $\beta$, which prohibits to resolve very small frequency. } \label{fig:Gtau_A0}
\end{figure}

It is also possible to do the DMFT self-consistent in the real time domain, but in the DMFT self-consistent loop we adopt the imaginary time Green's function which is more stable. And since all of the evolution operators are real, this choice would reduce the numerical effort. Once convergence is achieved, a real time evolution could be redone to get the retarded Green's function $G^{r}(t)$ and the spectral function could be extracted from it with appropriate FT method. 

\section{Summary and Outlooks}

Here we discuss various sources of error and computational effort of
the tVMPS solver. First due to the cutoff of the matrix size in the
MPS representation, there are errors in the ground state and the
evolution procedure. Also there are Trotter decomposition errors
from the discretization of the evolution operators. Finally, errors
also come from the mapping from the hybridization function to the
chain Hamiltonian. The latter two errors can be easily reduced by
adopting higher order Trotter decomposition and by increasing the
number of the fitting parameters, \textit{i.e.} the length of the
chain. However, the first error is more severe and prohibit long
time evolution. It is known that the algorithm will breakdown at
runaway time, which approximately shows the logarithmic dependence
on $\chi$. At the runaway time the entanglement of the states
increases to a value which could not be sufficiently represented by
the given finite truncation dimension. The computational effort of
the tVMPS solver is mainly determined by the truncation dimension
$\chi$ and is insensitive to the Hamiltonian parameters of the
problem.

We then summarize the limitations due to present implementation of
the algorithm, and point out possible solutions. First, the
algorithm does not have a good resolution on the Kondo resonant
peak, \textit{i.e.} it does not pin down at the noninteracting value
as $U$ increases. This could be cured by using the logarithmic
discretization in the bath, and \cite{Silva:2008p7061,HeidrichMeisner:2009p12615} has attained
some success in using this technique to resolve Kondo effect in the
transport properties. Second, to reduce computational effort we
limit chain length only slightly larger than that accessible to
exact diagonalization, but by using conserved quantity such as total
particle number, magnetization or even non-abelian $SU(2)$ symmetry
of SIAM, much larger system size is accessible within the same
computational time. Third, accurate determination of low temperature
Green's function needs long time evolution without hitting the
runaway time. Whether the tVMPS could produce result that competes
to the Monte Carlo method should be answered by future
investigation.

To conclude, we have shown that the tVMPS approach could work at zero temperature as well as at finite temperatures. By the post-selfconsistent real time evolution, one could also get the real time dynamics, from which the real frequency can be extracted. These merits make it a promising solver for DMFT to investigate the low temperature properties of the strongly correlated systems.

As an outlook, the time-dependent VMPS solver reported here can be
generalized to complex impurities, for example, the multi-orbital
impurity problems, and the quantum cluster problems. Different from
the ED based solvers, the generalization of the present solver to
multi-band or clusters case does not encounter the exponential
increasing of states. Therefore tVMPS is a promising solver to work
within LDA+DMFT\cite{Kotliar:2006p5576, Held:2007p4809} to
investigate the realistic materials and a cluster solver works
within the cluster-DMFT formalisms. \cite{Maier:2005p5551}. The
Trotter errors can be avoided even by adopting different evolution
algorithm, \textit{i.e.}, the Lanczos dynamics
\cite{Manmana:2005p6002} and the time step targeting
\cite{Schollwoeck:2006p1926}. Different contract schemes (transverse
contract) in the time evolution are reported
\cite{Banuls:2009p6102}. New algorithm for the finite temperature
based on the minimally entangled typical quantum states is reported
in \cite{White:2009p6808} and can be used to replace the ancilla
approach. The present solver works in real time domain, and thus may
be useful in the non-equilibrium DMFT \cite{Freericks:2006p6866}. In
an independent direction, the impurity solver developed here could
also be used to study the time-dependent phenomena in the quantum
transport in nanodevices. \cite{Elzerman:2004p6878}
\cite{Anders:2005p6185}

\begin{acknowledgements}
The work is supported by NSFC. XCX is supported by US-DOE and NSF. The authors thank Ning-Hua Tong, Quan-Sheng Wu, Li Huang and Zi Cai for helpful discussions.
\end{acknowledgements}

\begin{appendix}

    \section{Finite Temperature Algorithm \label{sec:temp}}

\begin{figure}
        [tbp] \centering
        \includegraphics[height=3cm, width=8cm]{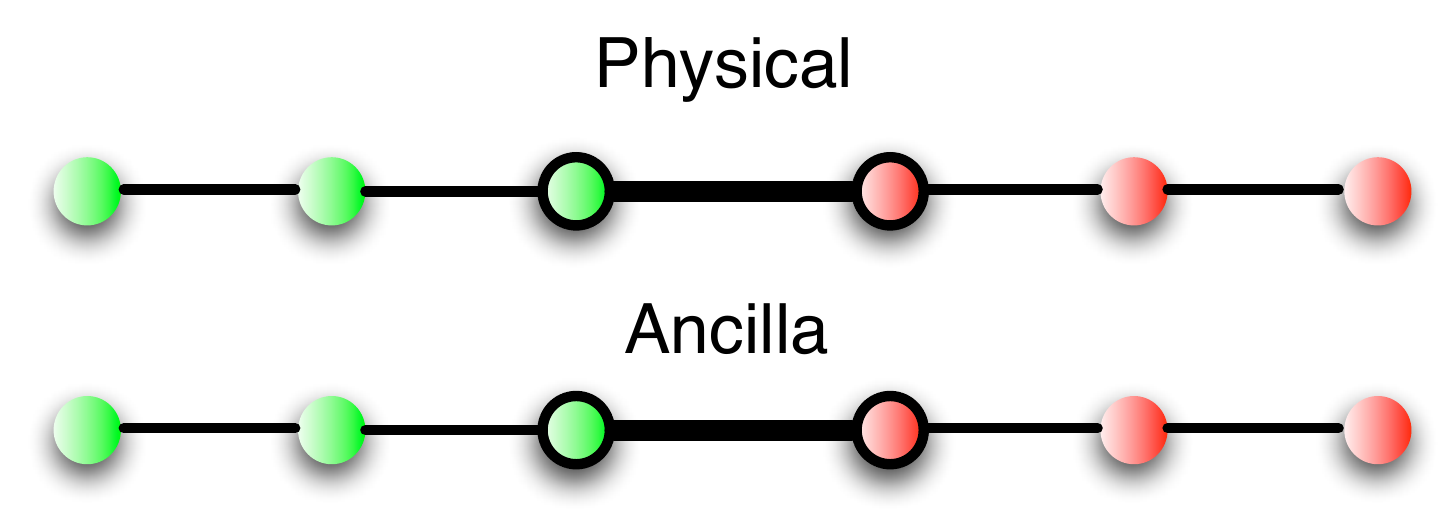} \caption{(Color online) Introduce the ancillary sites parallel to the SIAM chian. The ancillary sites act as the heat bath. When been traced out they give the thermodynamic averages. The presence of the ancillary site enlarges the local Hilbert space. The Hamlitonian only acts on the physical chain.} \label{fig:ancilla}
    \end{figure}

    In this section, we discuss the finite temperature algorithm based on the ancilla approach.\cite{Verstraete:2004p5961, Feiguin:2005p6426, Barthel:2009p6149} The ancilla approach replaces the mixed state needed by computing the thermodynamical averages by the pure state of an enlarged system. The enlarged system is constructed simply by the original physical system $\mathcal{H}$ and the identical copy of it, whose Hilbert space is denoted by $\mathcal{A}$. For the present case, the ancillary sites are added parallel to the SIAM chain, make the system resemble a ladder, and the Hamiltonian acts only on the physical sites, see Fig.\ref{fig:ancilla}. One first prepares the states,
    \begin{eqnarray}
        |\psi_0\rangle=\bigotimes_{i}\left(\sum_{\sigma_{i}}|\sigma_{i} \sigma_{i}\rangle\right)
    \end{eqnarray}
    where the first (second) $\sigma_{i}\in \mathcal{H}(\mathcal{A})$ denotes the state of the $i$-th physical (ancillary) site. By the purification process one gets $|\psi_{\beta}\rangle=e^{-\beta H/2}|\psi_0\rangle$, from which one could get the finite temperature density operator $\rho_{\beta}=\Tr_{\mathcal{A}}|\psi_{\beta}\rangle\langle\psi_{\beta}|$ and the partition function $Z=\langle\psi_{\beta}|\psi_{\beta}\rangle$. The finite temperature correlation function can be evaluated as $\Tr_{\mathcal{H}} (\rho_{\beta}c(\tau) c^{\dagger}) = \langle\psi_{\beta}|e^{H\tau} c e^{-H\tau} c^{\dagger} |\psi_{\beta}\rangle $. So once one gets $|\psi_{\beta}\rangle$, one could apply the evolution algorithm similar to the Sec.\ref{sec:evolution} to compute the Green's function in the imaginary as well as real times.

\begin{figure}
        [tbp] \centering
        \includegraphics[height=6cm, width=8cm]{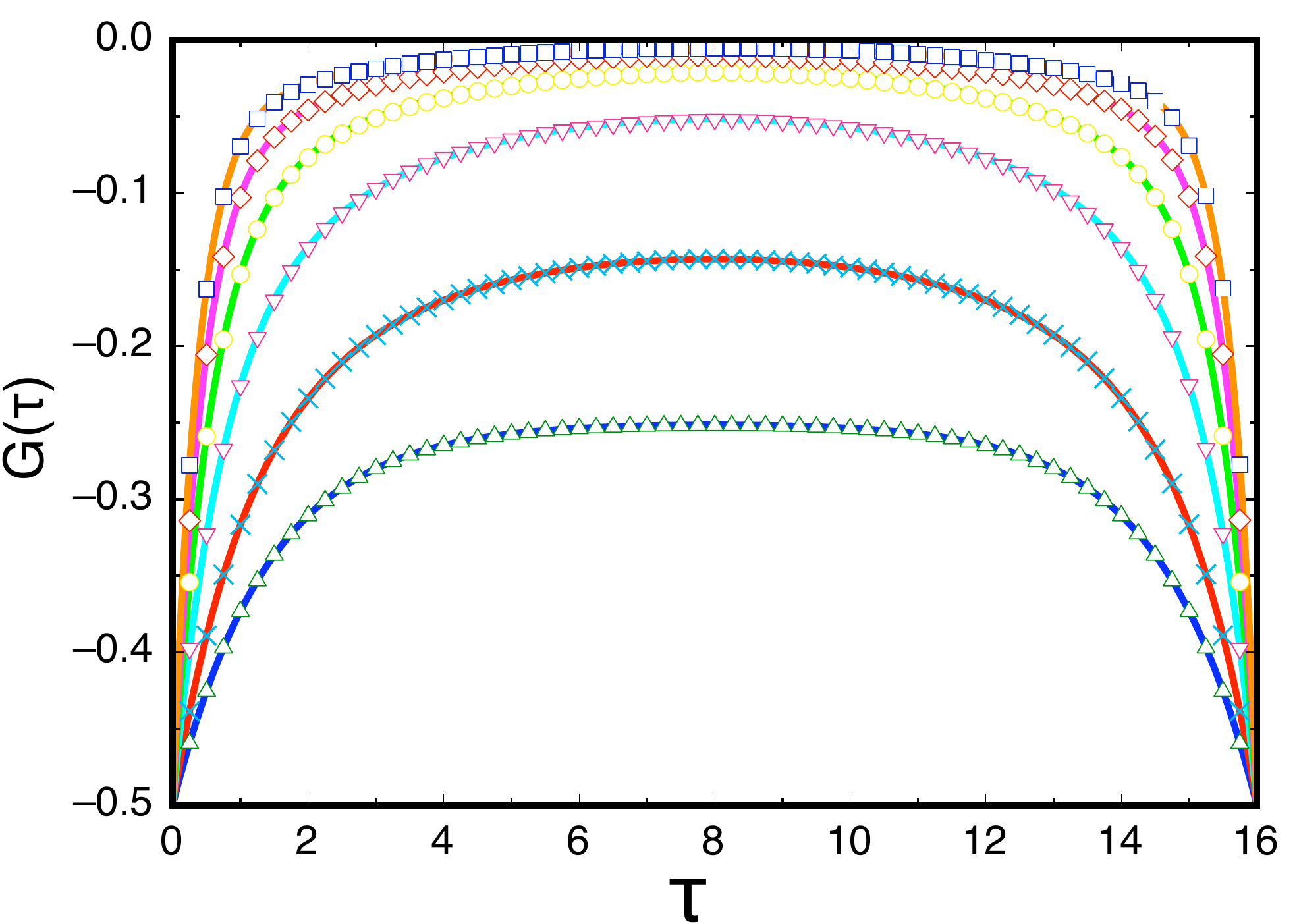} \caption{(Color online) The $\beta=16$ imaginary time Green's function for a $4$-site SIAM chain calculated by the tVMPS approach (lines) compared with the HF-QMC results (hollow dots). $\varepsilon_n=0, \gamma_n=0.5,U=0,1,2,3,4,5$ from bottom to top. $\tau>\beta/2$ data are calculated from the $-\beta/2<\tau<0$ data by $G(\tau)=-G(\tau-\beta)$ to avoid long time evolutions.} \label{fig:comp_qmc}
    \end{figure}

    The imaginary time Green's function $G(\tau)$ with $\tau=0$ to $\beta$ is computed from the evolution algorithm. For the particle-hole symmetrical case, one only needs half of the data ($\tau<\beta/2$) due to the symmetric property $G(\tau)=G(\beta-\tau)$. For general cases, we evolute $G(\tau)$ from $0$ to $\pm\beta/2$ and then use $G(\beta/2<\tau<\beta)=-G(-\beta/2<\tau-\beta<0)$ to restore the imaginary time Green's function from $0$ to $\beta$. This approach reduces the computational efforts and also the accumulated errors in the long time evolution. (The ending point is conserved since in the present approach it is equal to $\langle\psi_{0}|c e^{-\beta H/2}c^{\dagger}e^{-\beta H/2}|\psi_0\rangle$ ) To compute the Green's function at low but finite temperatures, one needs to perform long time evolution, which is not stable due to the accumulation of the Trotter error and the truncation of the Hilbert space. The error is more severe in the finite temperature case, because of the presence of the exponent growth factor $e^{H\tau}$. This hinders the investigation of very low temperatures but can be circumvented by choosing alternative evolution algorithms \cite{Banuls:2009p6102, Manmana:2005p6002,White:2009p6808}. These possibilities demands further investigating. See Fig \ref{fig:comp_qmc} for a comparison of the finite-T Green's function for a SIAM chain from the tVMPS and the HF-QMC approach.

    There is an alternative way which does not group the physical and the ancillary sites into the supersite. This way reduces the size of the local Hilbert space, but the non next-nearest interaction will hinder the simple Trotter decomposition based time evolution algorithm. After all, the time evolution can be done by the Lanczos dynamic or the time step targeting which does not assume the next-nearest interactions.

\end{appendix}


\end{document}